\documentclass[atmosphere,article,submit,moreauthors,pdftex,12pt,a4paper]{mdpi} 
\lastpage{x}
\doinum{10.3390/------}
\pubvolume{xx}
\pubyear{2011}
\history{Received: xx / Accepted: xx / Published: xx}

\pdfoutput=1
\usepackage{amssymb,amsmath}
\usepackage[utf8]{inputenc}
\usepackage[T1]{fontenc}
\usepackage{multirow}
\usepackage{longtable}
\Title{On the Fractal Mechanism of Interrelation Between the Genesis, Size and Composition of Atmospheric Particulate Matters in Different Regions of the Earth}

\Author{Vitaliy D. Rusov $^{1,\star}$, Radomir Ilic $^{2}$, Radojko  R. Jacimovic $^{2}$, Vladimir N. Pavlovich $^3$, Yuriy A. Bondarchuk $^1$,  Vladimir N. Vaschenko $^4$, Tatiana N. Zelentsova $^1$,  Margarita E. Beglaryan $^1$, Elena P. Linnik $^1$, Vladimir P. Smolyar$^1$, Sergey I. Kosenko $^{1}$ and Alla A. Gudyma $^4$}

\address{%
$^{1}$ Odessa National Polytechnic University, Shevchenko av. 1, 65044 Odessa, Ukraine\\
$^{2}$ J. Stefan Institute, Jamova 39, 1000 Ljubljana, Slovenia\\
$^{3}$ Institute for Nuclear Research, Pr. Nauki 47, 03028 Kyiv, Ukraine\\
$^{4}$ State Ecological Academy for Postgraduate Education and Management, Uritskogo Str. 35, 03035 Kyiv, Ukraine}
\corres{E-Mail: siiis@te.net.ua, Tel.: +38-048-2641672; Fax: ++38-048-2641672.}

\abstract{Experimental data from the National Air Surveillance Network of Japan from 1974 to 1996 and from independent measurements performed simultaneously in the regions of Ljubljana (Slovenia), Odessa (Ukraine) and the Ukrainian “Academician Vernadsky” Antarctic station (64$^\circ$15W; 65$^\circ$15S), where the air elemental composition was determined by the standard method of atmospheric particulate matter (PM) collection on nucleopore filters and subsequent neutron activation analysis, were analyzed. Comparative analysis of different pairs of atmospheric PM element concentration data sets, measured in different regions of the Earth, revealed a stable linear (on a logarithmic scale) correlation, showing a power law increase of every atmospheric PM element mass and simultaneously the cause of this increase – fractal nature of atmospheric PM genesis. Within the framework of multifractal geometry we show that the mass (volume) distribution of atmospheric PM elemental components is a log normal distribution, which on a logarithmic scale with respect to the random variable (elemental component mass) is identical to normal distribution. This means that the parameters of two-dimensional normal distribution with respect to corresponding atmospheric PM-multifractal elemental components measured in different regions, are a priory connected by equations of direct and inverse linear regression, and the experimental manifestation of this fact is the linear correlation between the concentrations of the same elemental components in different sets of experimental atmospheric PM data.}


\begin{document}

\makeatletter
  \renewcommand{\thesubsubsection}{\arabic{section}.\arabic{subsection}.\arabic{subsubsection}.}
  \renewcommand{\subsubsection}{ %
      \@startsection{subsubsection}{3}{0pt}{12pt}{12pt}{\normalsize\itshape}}
\makeatother
  

\section{Introduction}
Analysis of the concentrations of elements characteristics of the terrestrial crust, anthropogenic emissions and marine elements used in monitoring of the levels of atmospheric aerosol contamination indicates that these levels at the two extremes of: (i) the Antarctic (South Pole \cite{ref-1}), and (ii) the global constituent of atmospheric contamination measured at continental background stations \cite{ref-2} display similar patterns. A distinction between them is, however, evident and consists in that the mean element concentrations in the atmosphere over continental background stations ($C_{CB}$) located in different regions of the Earth exceed the corresponding concentrations at the South Pole ($C_{SP}$) by some $20-10^3$ times.

Comparing the concentrations of a given element {\it i} in atmospheric aerosol from samples $\left\lbrace C_{SP,i} \right\rbrace$ and $\left\lbrace C_{CB,i} \right\rbrace$, it became evident that the dependence of the mean concentrations of any particular element from the sample $\left\lbrace C_{SP,i} \right\rbrace$ or $\left\lbrace C_{CB,i} \right\rbrace$ on a logarithmic scale is described by a linear one, with good precision for any of the elements from a given pair of sampling station:

\begin{equation}
\label{eq1}
  \ln{C_{CB}^i} = a_{CB-SP} + b_{CB-SP} \ln{C_{SP}^i},
\end{equation}

\begin{equation}
\label{eq2}
  b_{CB-SP}  \approx 1,
\end{equation}

\noindent where $ a_{CB-SP} $ and $b_{CB-SP}$ are the intercept and slope (regression coefficient) of the regression line, respectively.

This unique and rather unexpected result was first established by Pushkin and Mikhailov \cite{ref-2}. It is noteworthy, according to \cite{ref-2}, that the reason for the large enrichment of atmospheric aerosol with those elements which are exceptions to the linear dependence, is related mostly either to the anthropogenic contribution produced by extensive technological activity, e.g. the toxic elements (Sb, Pb, Zn, Cd, As, Hg), or to the nearby sea or ocean or sea as a powerful source of marine aerosol components (Na, I, Br, Se, S, Hg).

However, our numerous experimental data specify persistently that the Pushkin-Mikhailov dependence (\ref{eq1}) in actual fact is the particular case (at $b_{12}=1$) of more general of linear regression equation 

\begin{equation}
  \label{eq3}
  \ln \left( C_{1i} / \rho_i \right) = a_{12} + b_{12} \ln{ \left( C_{2i} / \rho_i \right)}
\end{equation}

\noindent where $C_i$ and $ \rho_i $ are the concentration and specific density of i-th isotope component in atmospheric PM measured in different regions (the indexes 1 and 2) of the Earth.

It is obvious, if we will be able to prove that the linear relation (\ref{eq3}) in element concentrations between the above mentioned samples reflects the more general fundamental dependence, it can be used for the theoretical and experimental comparison of atmospheric PM independently of the given region of the Earth. Moreover, the linear relation (\ref{eq3}) can become a good indicator of the elements defining the level of atmospheric anthropogenic pollution, and thereby to become the basis of method for determining a pure air standard or, to put it otherwise, the standard of the natural level of atmospheric pollution of different suburban zones. This is also indicated by the power law character of Eq. (\ref{eq3}), reflecting the fact that the total genesis of non-anthropological (i.e natural) atmospheric aerosols does not depend upon the geography of their origin and is of a fractal nature.

In our opinion, this does not contradict the existing concepts of microphysics of aerosol creation and evolution \cite{ref-3}, if we consider the fractal structure of secondary ($D_p > 1\mu m$) aerosols as structures formed on the prime inoculating centres, ($D_p < 1\mu m$). Such a division of aerosols into two classes – primary and secondary \cite{ref-3} – is very important since it plays the key role for understanding of the fractal mechanism of secondary aerosol formation, which show scaling structure with well-defined typical scales during aggregation on inoculating centres (primary aerosols) \cite{ref-4}.

The objective of this work was twofold: (i) to prove reliably the linear validity of Eq. (\ref{eq3}) through independent measurements with good statistics performed at different latitudes and (ii) to substantiate theoretically and to expose the fractal mechanism of interrelation between the genesis, size and composition of atmospheric PM measured in different regions of the Earth, in particular in the vicinity of Odessa (Ukraine), Ljubljana (Slovenia) and the Ukrainian Antarctic station “Academician Vernadsky” (64$^\circ$15W; 65$^\circ$15S).


\section{The linear regression equation and experimental data of National Air Surveillance Network of Japan}

In this study, experimental data \cite{ref-5} from the National Air Surveillance Network (NASN) of Japan for selected crustal elements (Al, Ca, Fe, Mn, Sc and Ti), anthropogenic elements (As, Cu, Cr, Ni, Pb, V and Zn) and a marine element (Na) in atmospheric particulate matter obtained in Japan for 23 years from 1974 to 1996 were evaluated.  NASN operated 16 sampling stations (Nopporo, Sapporo, Nonotake, Sendai, Niigata, Tokyo, Kawasaki, Nagoya, Kyoto-Hachiman, Osaka, Amagasaki, Kurashiki, Matsue, Ube, Chikugo-Ogori and Ohmuta) in Japan, at which atmospheric PM were regularly collected every month by a low volume air sampler and analyzed by neutron activation analysis (NAA) and X-ray fluorescence (XRF). During the evaluation, the annual average concentration of each element based on 12 monthly averaged data between April (beginning of financial year in Japan) and March was taken from NASN data reports. The long-term (23 years) average concentrations were determined from the annual average concentration of each element.

Analysis of the NASN data \cite{ref-5} shows that the highest average concentrations were observed in Kawasaki (Fe, Ti, Mn, Cu, Ni and V), Osaka (Na, Cr, Pb and Zn), Ohmuta (Ca) and Niigata (As), respectively. These cities are either industrial or large cities of Japan. Conversely, the lowest average concentrations were noticed in Nonotake (Al, Ca, Fe, Ti, Cu, Cr, Ni, V and Zn) and Nopporo (Mn, As and Pb), as expected \cite{ref-5}. On the basis of these results, Nonotake and Nopporo were selected as the baseline-remote area in Japan.

\begin{figure}[H]
  \centering
  \includegraphics[width=90mm]{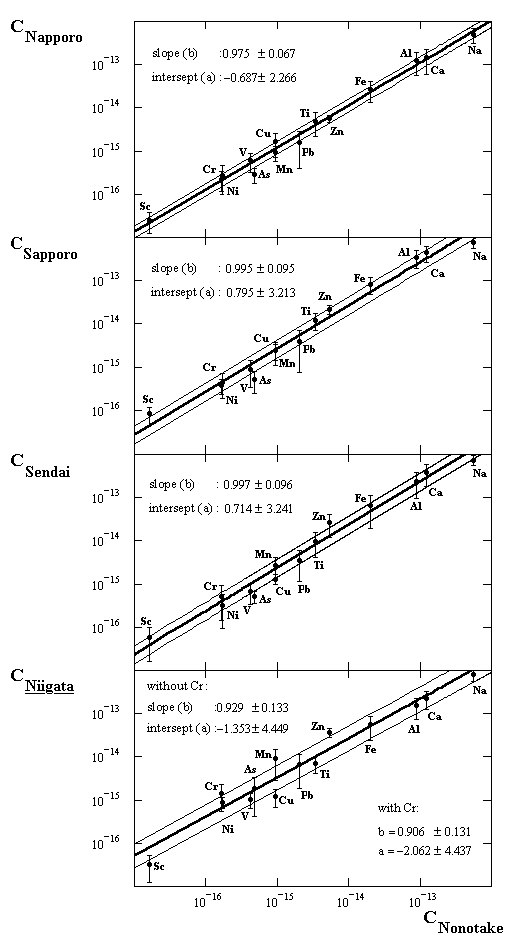}
  \caption{Relation between annual concentrations of elements in atmospheric particulate matter over Japan and the same data obtained in the region of Nonotake (data from NASN of Japan \protect\cite{ref-5}). In some cases (see text) the concentration of anthropogenic element \emph{Cr} was excluded from the data. The underlined cities are large industrial centres in the Japan \protect\cite{ref-5}.}
  \label{fig1a}
\end{figure}

A simple model of linear regression, in which the evaluations were made by the least squares method \cite{ref-6, ref-7}, was used to build the linear dependence described by Eq. (\ref{eq3}). The results of NASN data presented on a logarithmic scale relative to the data of Nonotake $ \left\lbrace C_{Nonotake}^i \right\rbrace$ (see Figure \ref{fig1a}) show with high confidence the adequacy of experimental and theoretical dependence of Eq. (\ref{eq3}) type. As can be seen from Figure \ref{fig1a}, the Nonotake station was chosen as the baseline, where the lowest concentrations of crustal and anthropogenic elements and small variations in time (23 years) were observed \cite{ref-5}. The NASN data are presented in the form of “city – city” concentration dependences.

\begin{figure}[H]
  \centering
  \includegraphics[width=90mm]{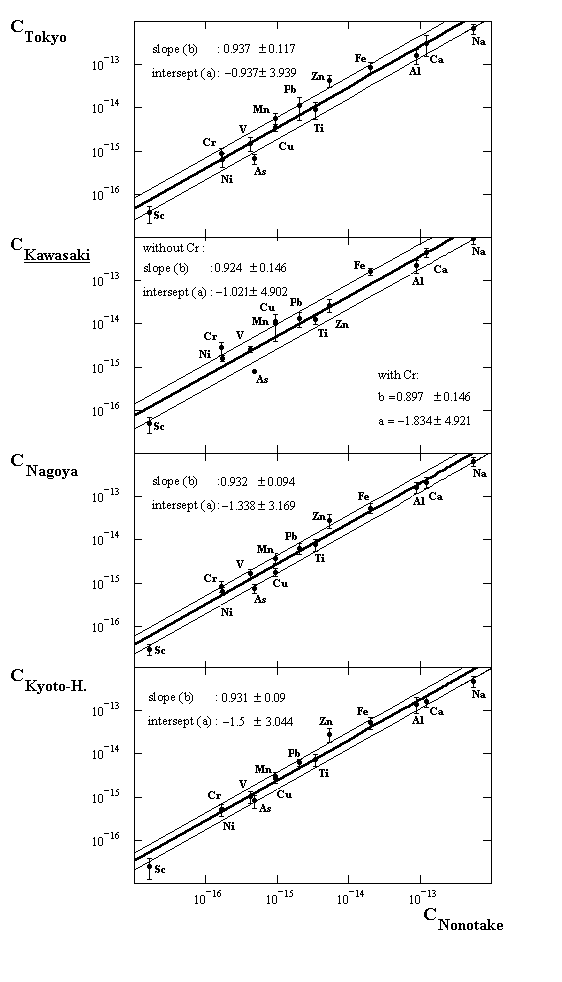}
  \label{fig1b}
\end{figure}

\begin{figure}[H]
  \centering
  \includegraphics[width=90mm]{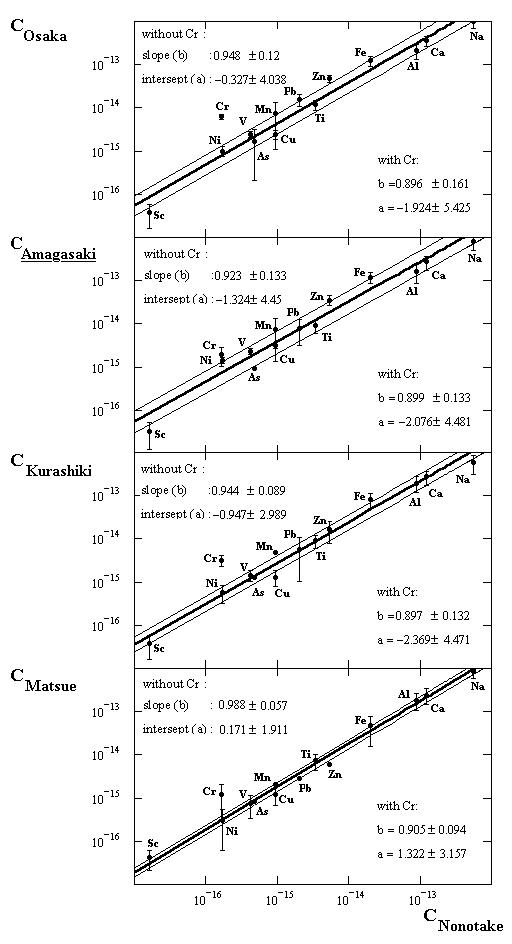} 
  \label{fig1c}
\end{figure}

\begin{figure}[H]
  \centering
  \includegraphics[width=90mm]{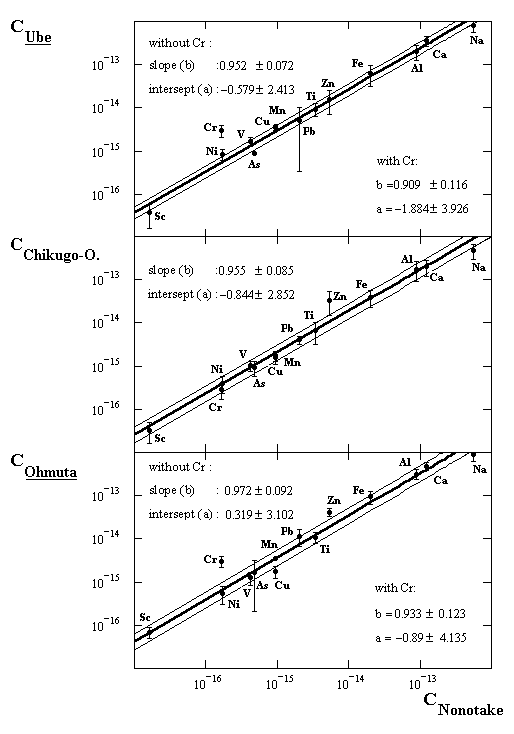}
  \label{fig1d}
\end{figure}

Analysis of concentration data for Japanese city atmospheric PM unambiguously shows that the \textit{i}-th element mass in atmospheric PM grows by power low, proving the assumption \cite{ref-4,ref-8,ref-9,ref-10} about the fractal nature of atmospheric PM genesis. 
\section{The linear regression equation and the composition of atmospheric aerosols in different regions of the Earth}

It is evident that in order to generalize the results from NASN data processing more widely, the validity of Eq. (\ref{eq3}) should be checked on the basis of atmospheric aerosol studies in performed independent experiments at different latitudes. For this reason such studies were performed in the regions of Odessa (Ukraine), Ljubljana (Slovenia) and the Ukrainian Academecian Vernadsky Antarctic station (64$^\circ$15W; 65$^\circ$15S). The determination of the element composition of the atmospheric air in these experiments was performed by the traditional method based on collection of atmospheric aerosol particles on nucleopore filters with subsequent use of $k_0$-instrumental neutron activation analysis. Regression analysis was used for processing of the experimental data.

\subsection{Experimental}

\subsubsection{Collection of atmospheric aerosol particles on nucleopore filters}
\begin{figure}
\centering
\includegraphics[scale=1]{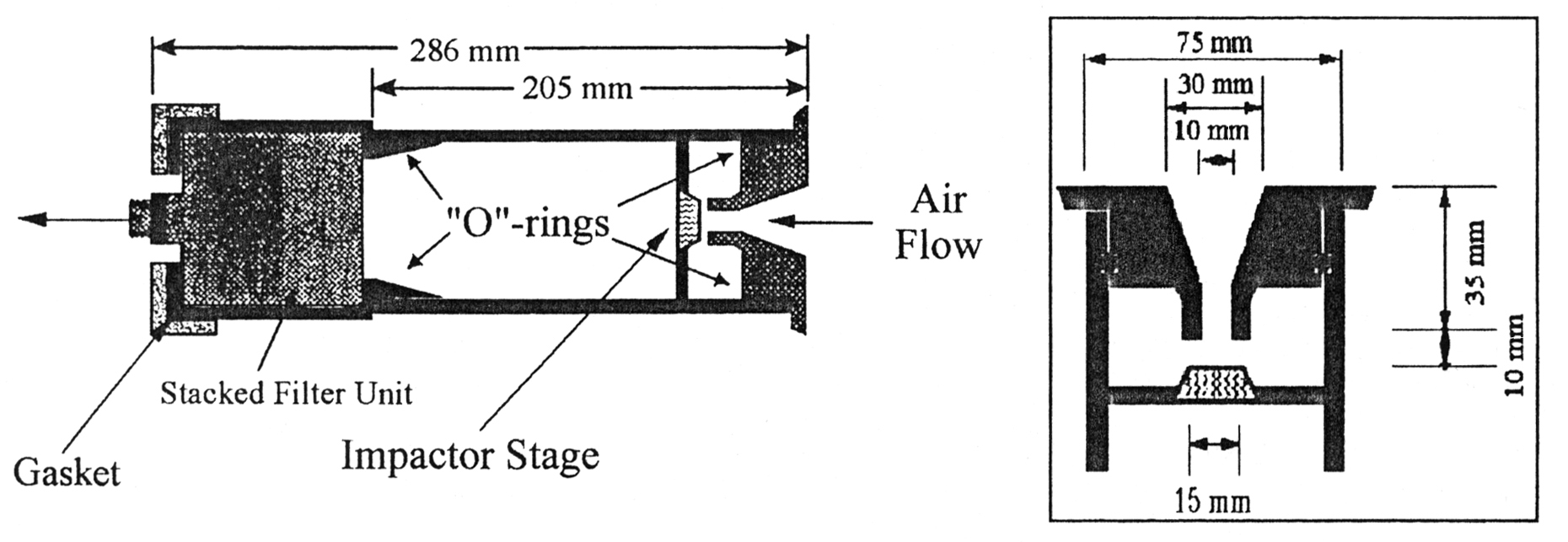} 
\caption{Schematic representation of the flow-chamber of the PM10 type device for air sampling.}
\label{fig2}
\end{figure}

For collection of atmospheric aerosol particles on filters, a device of the PM10 type was used with the Gent Stacked Filter Unit (SFU) interface \cite{ref-11, ref-12}. The main part of this device is a flow-chamber containing an impactor, the throughput capacity of which is equivalent to the action of a filter with an aerodynamic pore diameter of 10 $ \mu m $ and a 50\% aerosol particle collection efficiency based on mass, and an SFU interface designed by the Norwegian Institute for Air Research (NILU) and containing two filters (Nucleopore) each 47 mm in diameter, the first filter with a pore diameter of 8 $ \mu m$  and the second filter with 0.4 $ \mu m$ pore diameter. It was experimentally found \cite{ref-12} that such geometry (Figure \ref{fig2}) results in an aerosol collection efficiency of the first filter of approximately 50\%, whereas for the second filter this value was close to 100\% \cite{ref-12}. More detailed results of thorough testing of a similar device can be found in \cite{ref-12}.

\subsubsection{$k_0$-instrumental neutron activation analysis}
Airborne particulate matter (APM) loaded filters were pelleted with a manual press to a pellet of 5 mm diameter and each packed in a polyethylene ampoule, together with an Al-0.1\%Au IRMM-530 disk 6 mm in diameter and 0.2 mm thickness and irradiated for short irradiations (2-5 min) in the pneumatic tube (PT) of the 250 kW TRIGA Mark II reactor of the J. Stefan Institute at a thermal neutron flux of $3.5 \cdot 10^{12} cm^{-2} s^{-1}$, and for longer irradiations in the carousel facility (CF) at a thermal neutron flux of $1.1 \cdot 10^{12} cm^{-2} s^{-1}$ (irradiation time for each sample about 18-20 h). After irradiation, the sample and standard were transferred to clean 5 mL polyethylene mini scintillation vial for gamma ray measurement. To determine the ratio of the thermal to epithermal neutron flux ($f$) and the parameter $ \alpha $, which characterizes the degree of deviation of the epithermal neutron flux from the $1/E$-law, the cadmium ratio method for multi monitor was used \cite{ref-13}. It was found that $f = 32.9$ and $ \alpha = -0.026$ in the case of the PT channel, and $f = 28.7$ and $ \alpha = -0.015$ for the CF channel. These values were used in calculation of the concentrations of short- and long-lived nuclides.

$\gamma$-activity of irradiated samples were measured on two HPGe-detectors (ORTEC, USA) of 20 and 40\% measurement efficiency \cite{ref-13}. Experimental data obtained on these detectors were fed into and processed on EG\&G ORTEC Spectrum Master and Canberra S100 high-velocity multichannel analyzers, respectively. To calculate net peak areas, HYPERMET-PC V5.0 software was used \cite{ref-14}, whereas for evaluation of elemental concentrations in atmospheric aerosol particles, KAYZERO/SOLCOI$^{®}$ software was used \cite{ref-15}. More details of the $k_0$-instrumental neutron-activation analysis applied could be found in \cite{ref-13}.

\subsection{Comparative analysis of atmospheric PM composition in different regions of the Earth}

\begin{figure}
\centering
\includegraphics[height=240mm]{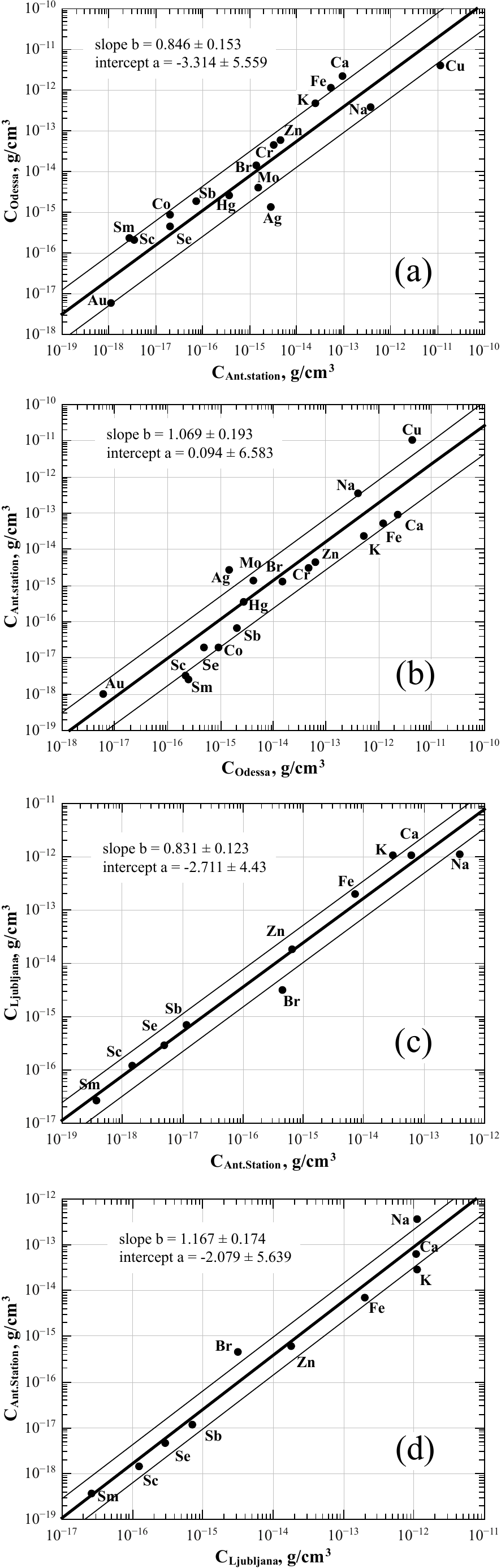} 
\caption{Lines of direct regression “Odessa - Antarctic station” (a), “Ljubljana - Antarctic Station” (c) and line of inverse regression “Antarctic Station - Odessa” (b), “Antarctic Station - Ljubljana” (d).}
\label{fig3}
\end{figure}

The results of presenting the atmospheric PM concentration values of Ljubljana $ \left\lbrace C_{Ljubljana}^i \right\rbrace$  and Odessa $ \left\lbrace C_{Odessa}^i \right\rbrace$ on a logarithmic scale relative to the similar data from Academician Vernadsky station $ \left\lbrace C_{Ant.station}^i \right\rbrace$ demonstrate with high reliability that the correlation coefficient r is approximately equal to unity both for the direct and reverse regression lines “Odessa - Antarctic station”, “Ljubljana - Antarctic station” (Figure \ref{fig3}):

\begin{equation}
  \label{eq4}
 r= \left[ b_{12} b_{21} \right]^{1/2} \approx 1,
\end{equation}

\noindent where $b_{12}$ and $b_{21}$ are the slopes of direct and reverse regression lines (Eq. (\ref{eq1})) for corresponding pairs.

\begin{figure}
\centering
\includegraphics[width=90mm]{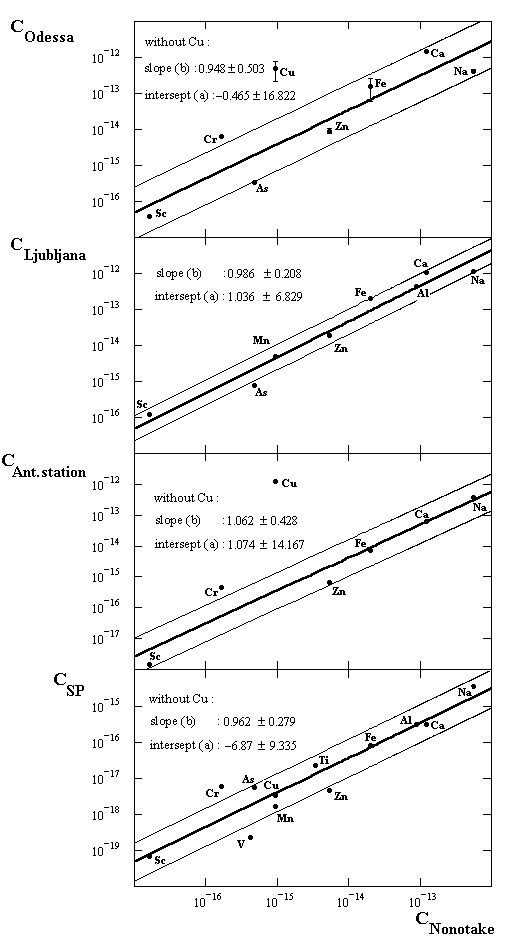} 
\caption{Relationship between atmospheric PM elemental concentrations measured in the regions Odessa (Ukraine), Ljubljana (Slovenia), Vernadsky station (64$^\circ$15W; 65$^\circ$15S), South Pole \protect\cite{ref-1} and the same data measured in the region of Nonotake \protect\cite{ref-5}.}
\label{fig4}
\end{figure}

\begin{figure}
\centering
\includegraphics[height=230mm]{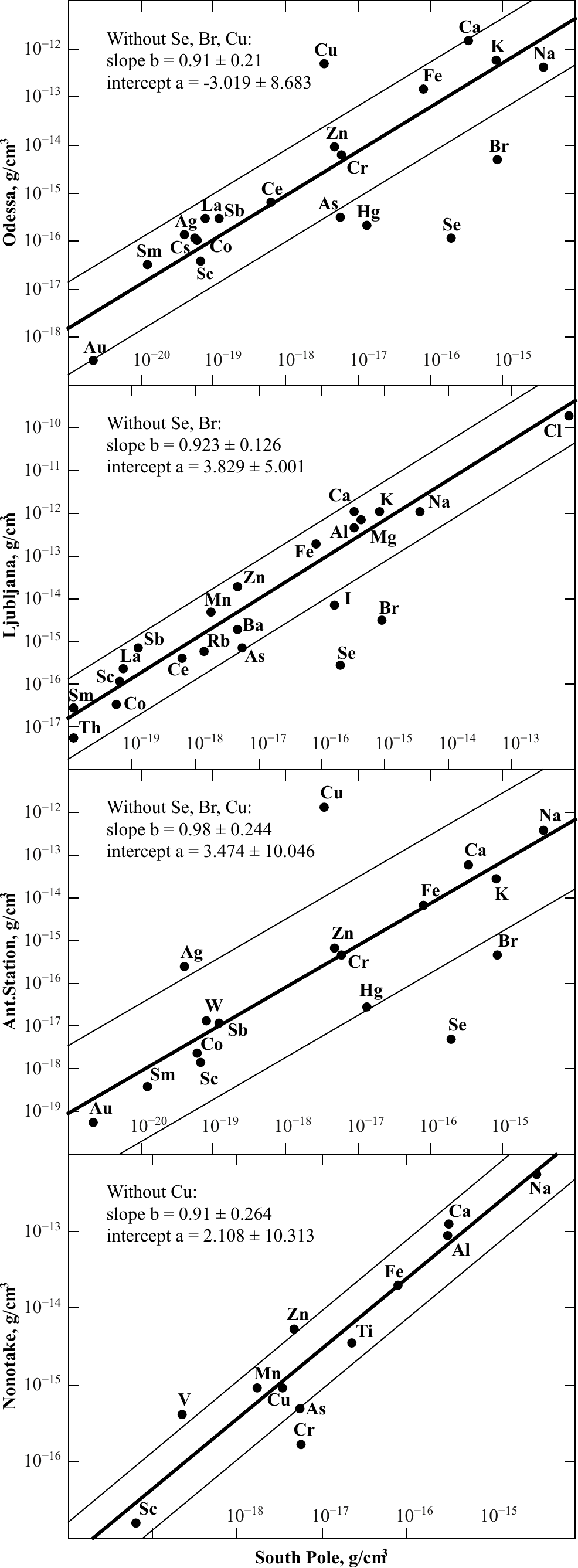} 
\caption{Relationship between atmospheric PM elemental concentrations measured in the regions Odessa (Ukraine), Ljubljana (Slovenia), Vernadsky station (64$^\circ$15W; 65$^\circ$15S), Nonotake \protect\cite{ref-5} and the same data measured in the region of the South Pole \protect\cite{ref-1}.}
\label{fig5}
\end{figure}

Figures \ref{fig4} and \ref{fig5} show the regression lines for daily normalized average concentrations of crustal, anthropogenic and marine elements (Table \ref{tab1}), measured on March 2002 in the regions of Odessa (Ukraine), the Ukrainian Antarctic station ($^\circ$15W; 65$^\circ$15S) and Ljubljana (Slovenia) relative to the same data obtained in Nonotake \cite{ref-5} (Figure \ref{fig4}) and the South Pole \cite{ref-1} (Figure \ref{fig5}).

  \begin{center} 
      \begin{longtable}{|c|c|c|c|c|c|c|}  
  \caption{\label{tab1} Chemical element composition in atmospheric PM. Symbols $C_{SP}$, $C_{CB}$, $C_{Nonotake}$ denote concentrations at the South Pole, continental background stations and Nonotake-city, respectively. Measured concentrations in the vicinity of the Ukrainian Antarctic Station, Odessa and Ljubljana are denoted by $C_{Antarctica}$, $C_{Odessa}$ and $C_{Ljubljana}$}\\
  
      \hline
      \multirow{2}{*}{Element} & $C_{SP}$ & $C_{CB}$ & $C_{Nonotake}$ & $C_{Antarctica}$ & $C_{Odessa}$ & $C_{Ljubljana}$ \\
      \cline{2-7}
      & \multicolumn{6}{|c|}{($ng / m^3$)} \\
      \hline
      S & 49 & - & - & - & - & - \\
      \hline
      Si & - & - & - & - & - & - \\
      \hline
      Cl & 2.4 & 90 & - & - & - & 528.2 \\
      \hline
      Al & 0.82 & 1.2 $\times$ 10$^3$ & 237.4 & - & - & 1152 \\
      \hline
      Ca & 0.49 & - & 185.8 & 93 & 2230 & 1630 \\
      \hline
      Fe & 0.62 & 1 $\times$ 10 $^2$ & 157.6 & 53.4 & 1201 & 1532 \\
      \hline
      Mg & 0.72 & - & - & - & - & 1264 \\
      \hline
      K & 0.68 & - & - & 24.4 & 502.8 & 918 \\
      \hline
      Na & 3.3 & 1.4 $\times$ 10 $^2$ & 538.8 & 361.95 & 393.8 & 1046 \\
      \hline
      Pb & - & 10 & 23.2 & - & - & - \\
      \hline
      Zn & 3.3 $\times$ 10$^{-2}$ & 10 & 38.3 & 4.48 & 62.6 & 127.2 \\
      \hline
      Ti & 0.1 & - & 15.4 & - & - & - \\
      \hline      
      F & - & - & - & - & - & - \\
      \hline      
      Br & 2.6 & 4 & - & 1.34 & 14.81 & 9.62 \\
      \hline      
      Cu & 2 $\times$ 10 $^{-2}$ & 3 & 8.20 & 11050  & 4240 & - \\
      \hline      
      Mn & 1.2 $\times$ 10 $^{-2}$ & 3 & 6.61 & - & - & 34.7 \\
      \hline      
      Ni & - & 1 & 1.48 & - & - & - \\
      \hline      
      Ba & 1.6 $\times$ 10 $^{-2}$ & - & - & - & - & 6.91 \\
      \hline      
      V & 1.3 $\times$ 10 $^{-3}$ & 1 & 2.44 & - & - & 17.08 \\
      \hline      
      I & 0.74 & - & - & - & - & 36.24 \\
      \hline      
      Cr & 4 $\times$ 10 $^{-2}$ & 0.8 & 1.14 & 3.11 & 45.2 & - \\
      \hline      
      Sr & 5.2 $\times$ 10 $^{-2}$ & - & - & - & - & - \\
      \hline      
      As & 3.1 $\times$ 10 $^{-2}$ & 1 & 2.74 & - & 1.83 & 4.18 \\
      \hline      
      Rb & 2 $\times$ 10 $^{-3}$ & - & - & - & - & 0.92 \\
      \hline      
      Sb & 8 $\times$ 10 $^{-4}$ & 0.5 & - & 0.07 & 1.97 & 4.62 \\
      \hline      
      Cd & < 1.5 $\times$ 10 $^{-2}$ & 0.4 & - & - & - & - \\
      \hline      
      Mo & 49 & - & - & 1.46 & 4.08 & - \\
      \hline      
      Se & < 0.8 & 0.3 & - & 0.02 & 0.47 & 1.21 \\
      \hline      
      Ce & 4 $\times$ 10 $^{-3}$ & - & - & - & 4.22 & 2.72 \\
      \hline      
      Hg & 0.17 & 0.3 & - & 0.36 & 2.75 & - \\
      \hline      
      W & 1.5 $\times$ 10 $^{-3}$ & - & - & 0.24 & - & - \\
      \hline      
      La & 4.5 $\times$ 10 $^{-4}$ & - & - & - & 1.79 & 1.36 \\
      \hline      
      Ga & < 1 $\times$ 10 $^{-3}$ & - & - & - & - & - \\
      \hline      
      Co & 5 $\times$ 10 $^{-4}$ & 0.1 & - & 0.02 & 0.89 & 0.29 \\
      \hline      
      Ag & < 4 $\times$ 10 $^{-4}$ & - & - & 2.74 & 1.42 & - \\
      \hline      
      Cs & 1 $\times$ 10 $^{-4}$ & - & - & - & 0.09 & - \\
      \hline      
      Sc & 1.6 $\times$ 10 $^{-4}$ & 5 $\times$ 10 $^{-2}$ & 0.04 & 3.36 $\times$ 10 $^{-3}$ & 0.21 & 0.298 \\
      \hline      
      Th & 1.4 $\times$ 10 $^{-4}$ & - & - & - & 0.29 & 0.060 \\
      \hline      
      U & - & - & - & - & 0.07 & - \\
      \hline      
      Sm & 9 $\times$ 10 $^{-5}$ & - & - & 2.68 $\times$ 10 $^{-3}$ & 0.24 & 0.198 \\
      \hline      
      In & 5 $\times$ 10 $^{-5}$ & - & - & - & - & - \\
      \hline      
      Ta & 7 $\times$ 10 $^{-5}$ & - & - & - & - & - \\
      \hline      
      Hf & 6 $\times$ 10 $^{-5}$ & - & - & - & - & - \\
      \hline      
      Yb & < 0.05 & - & - & - & - & - \\
      \hline      
      Eu & 2 $\times$ 10 $^{-5}$ & - & - & - & - & - \\
      \hline      
      Au & 4 $\times$ 10 $^{-5}$ & - & - & 1.07 $\times$ 10 $^{-5}$ & 0.006 & - \\
      \hline      
      Lu & 6.7 $\times$ 10 $^{-6}$ & - & - & - & - & - \\
      \hline      
      \end{longtable}
  \end{center}

The choice of the atmospheric aerosol concentration values of the South Pole as a baseline, relative to which a dependence of the type given by Eq. (\ref{eq1}) was analyzed, was made for three reasons: (i) these data were obtained by the same technique and method as in section 3.1, and simultaneously expand the geographical comparison, (ii) the element spectrum that characterizes the atmosphere of South Pole covers a wide range of elements (see Table 1); and (iii) the South Pole has the purest atmosphere on Earth, making it a convenient basis for comparative analysis.

\begin{figure}
\centering
\includegraphics[height=230mm]{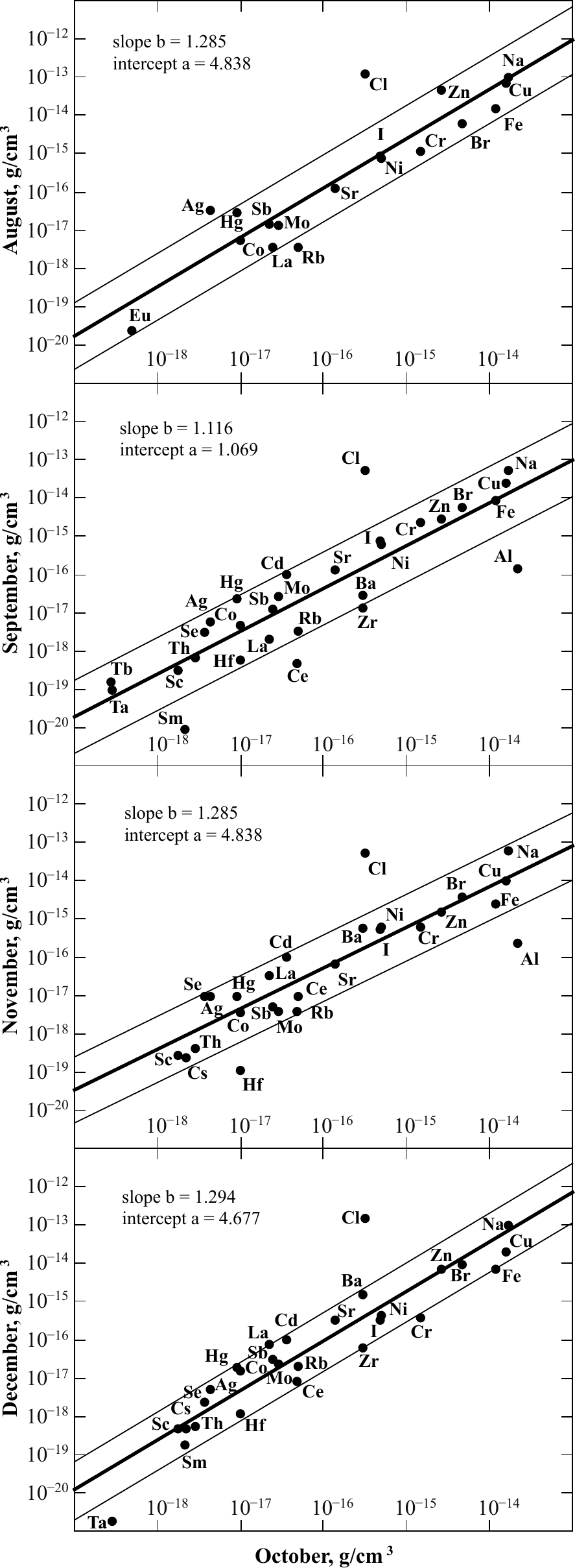} 
\caption{The regression lines for the normalized monthly average concentrations of atmospheric PM measured in the region of the Ukrainian Antarctic station “Academician Vernadsky” measured in August--December 2006 with respect to October 2006.}
\label{fig6}
\end{figure}

\begin{figure}
\centering
\includegraphics[width=90mm]{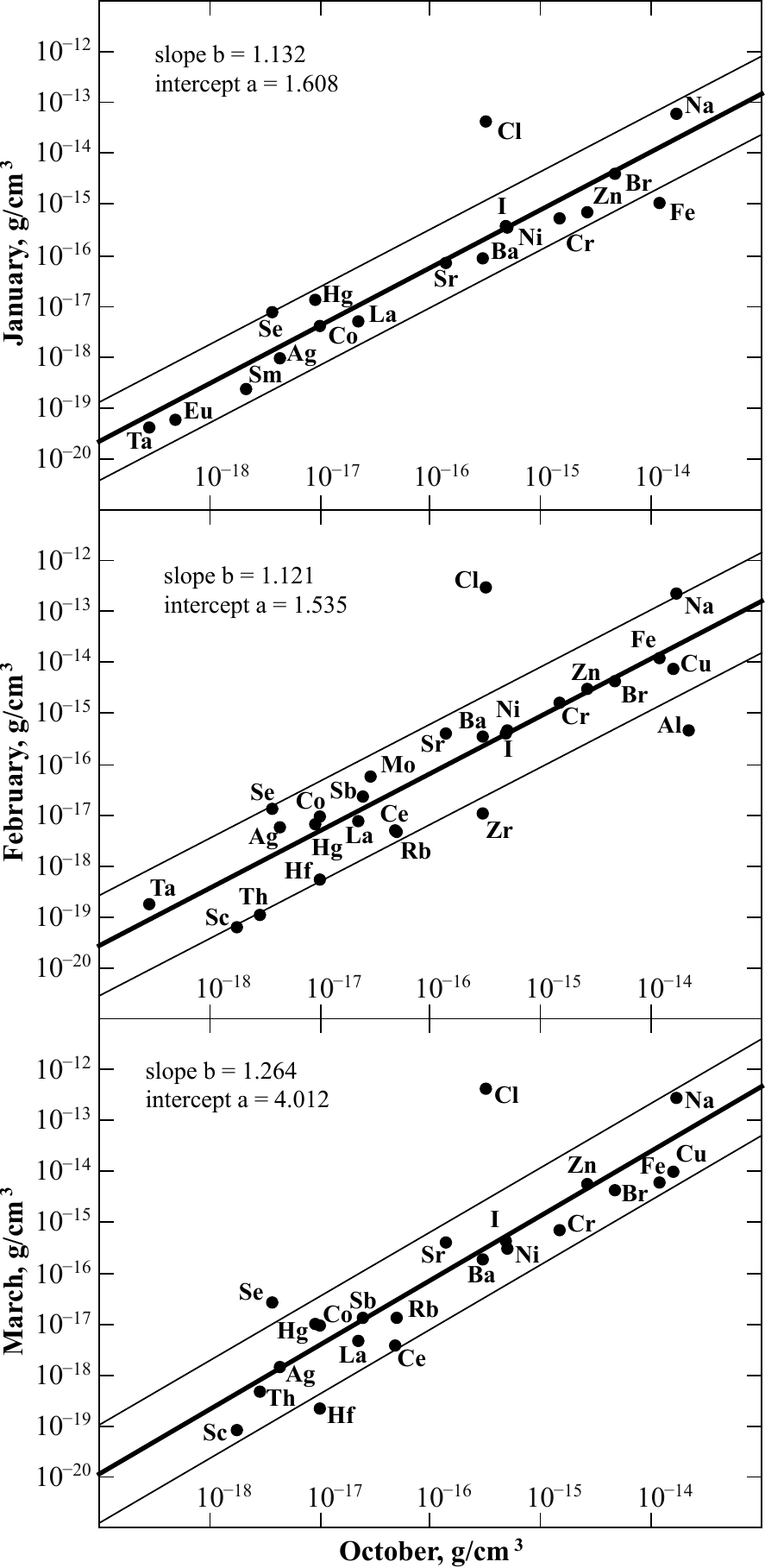} 
\caption{The regression lines for the normalized monthly average concentrations of atmospheric PM measured in the region of the Ukrainian Antarctic station “Academician Vernadsky” in January-March 2007 with respect to October 2006.}
\label{fig7}
\end{figure}

We present also the monthly normalized average concentrations of atmospheric PM measured over the period 2006-2007 in the region of the Ukrainian Antarctic station “Academician Vernadsky” (Figures \ref{fig6}, \ref{fig7}).

Comparative analysis of experimental sets of normalized concentrations of atmospheric aerosol elements measured in our experiments and independent experiments of the Japanese National Air Surveillance Network (NASN) shows a stable linear (on a logarithmic scale) dependence on different time scales (from average daily to annual). That points to a power law increase of every atmospheric PM element mass (volume) and simultaneously to the cause of this increase -- the fractal nature of atmospheric PM genesis.

In other words, stable fulfillment of the equality (\ref{eq4}) not only for the experimental data shown in Figures \ref{fig3}-\ref{fig7}, but also for any pairs of the  NASN data \cite{ref-5} (see Figure \ref{fig2}) unambiguously indicates that, on the one hand, the model of linear regression is satisfactory and, on the other hand, any of the analyzed samples $ \left\lbrace m_i \right\rbrace$, which describe the sequence of \textit{i}-th element partial concentrations in an aerosol, must to obey the Gauss distribution with respect to the random quantity $ \ln{p_i}$. Proof of these assertions for multifractal objects is presented below.


\section{The spectrum of multifractal dimensions and log normal mass distribution of secondary aerosol elements}

A detailed analysis of Figs. 1, 3-5, where the linear regressions for different pairs of experimental samples of element concentrations in atmospheric PM measured at various latitudes are shown, allows us to draw a definite conclusion about the multifractal nature of PM. The basis of such a conclusion is the reliably observed linear dependence of Eq. (\ref{eq3}) type between the normalized concentrations $C_i$ of the same element \textit{i} in atmospheric PM in different regions of the Earth.

Thus, it is necessary to consider the atmospheric PM, which is the multicomponent (with respect to elements) system, as a nonhomogeneous fractal object, i.e., as a multifractal. At the same time, the spectrum of fractal dimensions $f( \alpha )$ and not a single dimension $ \alpha_0$ (which is equal to $  D_0 $ for a homogeneous fractal) is necessary for complete description of a nonhomogeneous fractal object. We will show below that the spectrum of fractal dimensions $  f(\alpha) $ of multifractal predetermines the log normal type of statistics or, in other words, the log normal type of mass distribution of multifractal i-th component. This is very important, because the representation of mass distribution of atmospheric PM as the log normal distribution is predicted within the framework of the self-preserving theory \cite{ref-20,ref-21} and is confirmed by numerous experiments at the same time \cite{ref-3}.

To explain the main idea of derivation we give the basic notions and definitions of the theory of multifractals. Let us consider a fractal object which occupies some bounded region £ of size $L$ in Euclidian space of dimension $d$. At some stage of its construction let it represents the set of $N >> 1$ points distributed somehow in this region. We divide this region £ into cubic cells of side $ \varepsilon << L$ and volume $\varepsilon^d$. We will take into consideration only the occupied cells, where at least one point is. Let $i$ be the number of occupied cell $i = 1, 2,….N(\varepsilon)$, where $N(\varepsilon)$ is the total number of occupied cells, which depends on the cell size $\varepsilon$ . Then in the case of a regular (homogeneous) fractal, according to the definition of fractal dimensions $D$, the total number of occupied cells $N(\varepsilon)$ at quite small $\varepsilon$ looks like

\begin{equation}
\label{eq5}
 N(\varepsilon) \approx \varepsilon_L^{-D} = L_\varepsilon^D .
\end{equation}

\noindent where $\varepsilon$ is the cell size in $L$ units, $L_\varepsilon$ is the size of fractal object in $\varepsilon$ units.

When a fractal is nonhomogeneous a situation becomes more complex, because, as was noted above, a multifractal is characterized by the spectrum of fractal dimensions $f(\alpha)$, i.e. by the set of probabilities $p_i$, which show the fractional population of cells $\varepsilon$ by which an initial set is covered. The less the cell size is, the less its population. For self-similarity sets the dependence $p_i$ on the cell size $\varepsilon$ is the power function

\begin{equation}
\label{eq6}
  p_i ( \varepsilon) = \frac{1}{N_i (\varepsilon)} \approx \varepsilon_{L}^{ \alpha_{i} } = L_{ \varepsilon }^{-\alpha_i}
\end{equation}

where $\alpha_i$ is a certain exponent which, generally speaking, is different for the different cells $i$. It is obvious, that for a regular (homogeneous) fractal all the indexes $\alpha_i$ in Eq. (\ref{eq6}) are identical and equal to the fractal dimension $D$.

We now pass on to probability distribution of the different values $\alpha_i$. Let $n(\alpha)d\alpha$ is the probability what $\alpha_i$ is in the interval $\left[\alpha, \alpha + d\alpha \right]$. In other words, $n(\alpha)d\alpha$ is the relative number of the cells $i$, which have the same measure $p_i$ as $\alpha_i$ in the interval $\left[\alpha, \alpha + d\alpha \right]$. According to (\ref{eq5}), this number is proportional to the total number of cells $N(\varepsilon) \approx \varepsilon_L^{-D}$ for a monofractal, since all of $\alpha_i$ are identical and equal to the fractal dimension $ D $.

However, this is not true for a multifractal. The different values of $\alpha_i$ occur with probability characterized by the different (depending on $\alpha$) values of the exponent $f(\alpha)$, and this probability inherently corresponds to a spectrum of fractal dimensions of the homogeneous subsets $\text{£}_\alpha$ of initial set £:

\begin{equation}
\label{eq7}
 n ( \alpha ) \approx \varepsilon_L^{ - f( \alpha)}.
\end{equation}

Thus, from here a term “multifractal” becomes clear. It can be understood as a hierarchical joining of the different but homogeneous fractal subsets $\text{£}_\alpha$ of initial set £, and each of these subsets has the own value of fractal dimension $f(\alpha)$.

Now we show how the function $f(\alpha)$ predetermines the log normal kind of mass (volume) distribution of the multifractal $i$-th component. To ease further description we represent expression (\ref{eq8}) in the following equivalent form:

\begin{equation}
\label{eq8}
 n ( \alpha) \approx \exp{ \left[ f( \alpha ) \ln{L_\varepsilon }  \right]}.
\end{equation}

It is not hard to show \cite{ref-19} that the single-mode function $f(\alpha)$ can be approximated by a parabola near its maximum at the value $\alpha_0$.

\begin{equation}
\label{eq9}
  f( \alpha ) \approx D_0 - \eta \cdot  ( \alpha - \alpha_0 )^2,
\end{equation}

\noindent where the curvature of parabola

\begin{equation}
\label{eq10}
 \eta = \frac{f'' (\alpha_0) }{2} = \frac{1}{ 2 \left[ 2 ( \alpha_0 - D_0 ) + D''_{q=0} \right]}
\end{equation}

\noindent is determined by the second derivative of function $f(\alpha)$ at a point $\alpha_0$. Due to a convexity of the function $f(\alpha)$ it is obvious, that the magnitude in square brackets must be always positive. The fact, that the last summand $D''_{q=0}$ in these brackets is numerically small and it can be neglected, will be grounded below.

At the large $L_\varepsilon$ the distribution $n(\alpha)$ (\ref{eq8}) with an allowance for (\ref{eq9}) takes on the form

\begin{equation}
\label{eq11}
 n ( \alpha ) \sim \exp{ \left[ D_0 \ln{L_\varepsilon} \frac{ \left( \alpha - \alpha_0 \right)^2 \ln{L_\varepsilon}}{4 \left( \alpha_0 - D_0 \right) } \right]}.
\end{equation}

Then, taking into account (\ref{eq5}), we obtain from (\ref{eq11}) the distribution function of random variable $p_i $

\begin{equation}
 n ( p_i )  \sim \ln{N_{D_0}} \cdot \exp{ \left[ - \frac{1}{4 \ln{\left( 1/p_0 N_{D_0} \right)} } \left( \ln{p_i} - \ln{p_0} \right)^2 \right]},
\end{equation}

\noindent which with consideration of normalization takes on the final form

\begin{equation}
\label{eq13}
P ( p_i ) =  \frac{1}{ \sqrt{2 \pi} \sigma } \exp{ \left( -\mu - \frac{1}{2} \sigma^2 \right) } \cdot \exp{ \left[ - \frac{1}{2 \sigma^2} \left( \ln{p_i}  -\mu \right)^2 \right]},
\end{equation}

\noindent where

\begin{equation}
 \mu = \ln{p_0}, ~~~ \sigma^2 = 2 \ln{\frac{1}{p_0 N_{D_0}}}.
\end{equation}

This is the so-called log normal (relative) mass $p_i$ distribution. It is possible to present the first moments of such a kind of distribution for random variable $p_i$  in the following form:

\begin{equation}
\label{eq15}
\left\langle p \right\rangle = \exp{ \left( \mu + \frac{3}{2} \sigma^2 \right)} = 1/ p_0^2 N_{D_0}^3 = L_\varepsilon^{2 \alpha_0 - 3 D_0 },
\end{equation}

\begin{equation}
\label{eq16}
\mbox{var}(p) = \exp{\left( 2 \mu \right)} \cdot \left[ \exp{\left(4 \sigma^2 \right)} - \exp{\left(3 \sigma^2 \right)} \right] = L_\varepsilon^{2 \left( 2 \alpha_0 - 3D_0 \right)} \left( L_\varepsilon^{2 \left( 2 \alpha_0 - D_0 \right) } - 1\right).
\end{equation}

At the same time, it is easy to show that the distribution (\ref{eq13}) for the random variable lnpi has the classical Gaussian form

\begin{equation}
\label{eq17}
P \left( \ln{p} \right) = \frac{1}{\sqrt{2 \pi \sigma^2}} \exp{ \left\lbrace - \frac{1}{2 \sigma^2} \left[ \ln{p} - \left\langle \ln{p} \right\rangle \right]^2 \right\rbrace},
\end{equation}

\noindent where the first moments of this distribution for the random variable $\ln{p_i}$ look like

\begin{equation}
\label{eq18}
  \left\langle \ln{p} \right\rangle = \mu + \sigma^2 = \left( \alpha_0 -2 D_0 \right) \ln{L_\varepsilon},
\end{equation}

\begin{equation}
\label{eq19}
\mbox{var} \left( \ln{p} \right) = \sigma^2 = 2 \left( \alpha_0 - D_0 \right) \ln{L_\varepsilon}.
\end{equation}

Thus, according to the known theorem of multidimensional normal distribution shape \cite{ref-22}, a normal law of plane distribution for the two-dimensional random variable $(p_{1i}, p_{2i})$ will be written down as

\begin{equation}
P \left( \ln{p_{1i}}, \ln{p_{2i}} \right) = \frac{1}{2 \pi \sigma_1 \sigma_2 \sqrt{1 - r^2}} \exp{ \left[ - \frac{1}{2 \left( 1 - r^2 \right)} \left( u^2 + v^2 - 2ruv \right) \right]},
\end{equation}

\noindent where

\begin{equation}
u = \frac{\ln{p_{1i}} - \left\langle \ln{p_{1i}} \right\rangle}{\sigma_1}, ~~~ v = \frac{\ln{p_{2i}} - \left\langle \ln{p_{2i}} \right\rangle}{ \sigma_2},
\end{equation}

\noindent $r=\mbox{cov}(\ln{p_{1i}},\ln{p_{2i}}) / \sigma_1\sigma_2$ is the correlation coefficient between $\ln{p_{1i}}$ and $\ln{p_{2i}}$.

Then, by virtue of well-known linear correlation theorem \cite{ref-6,ref-22} it is easy to show that $\ln{p_{1i}}$ and $\ln{p_{2i}}$ are connected by the linear correlation dependence, if the two-dimensional random variable $(\ln{p_{1i}},\ln{p_{2i}})$ is normally distributed. This means that the parameters of two-dimensional normal distribution of the random values $p_{1i}$ and $p_{2i}$ for the $i$-th component in one aerosol particle, which are measured in different regions of the Earth (the indexes 1 and 2), are connected by the equations of direct linear regression:

\begin{equation}
\ln{p_{1i}} - \left\langle \ln{p_{1i}} \right\rangle = r \frac{\sigma_{1i}}{\sigma_{2i}} \left[ \ln{p_{2i}} - \left\langle \ln{p_{2i}} \right\rangle \right]
\end{equation}

\noindent and inverse linear regression

\begin{equation}
\ln{p_{2i}} - \left\langle \ln{p_{2i}} \right\rangle = r \frac{\sigma_{2i}}{\sigma_{1i}} \left[ \ln{p_{1i}} - \left\langle \ln{p_{1i}} \right\rangle \right],
\end{equation}

\noindent where  $i=1,…, N_p$ is the component number.

Taking into consideration that we measure experimentally the total concentration $C_i$ of the $i$-th component in the unit volume of atmosphere, the partial concentration $m_i$ of the $i$-th component in one aerosol particle measured in different regions of the Earth (the indexes 1 and 2) looks like

\begin{equation}
m_{1i} = C_{1i} / n_1, ~~~ m_{2i} = C_{2i} / n_2,
\end{equation}

\noindent where $n_1$ and $n_2$ are the number of inoculating centers, whose role play the primary aerosols $(D_p< 1 \mu m)$.

Here it is necessary to make important digression concerning the choice of quantitative measure for description of fractal structures. According to Feder \cite{ref-18}, determination of appropriate probabilities corresponding to the chosen measure is the main difficulty. In other words, if choice of measure determines the search procedure of probabilities $ \left\lbrace p_i \right\rbrace$, which describe the increment of the chosen measure for given level of resolution $\varepsilon$, then the probabilities themselves predetermine, in its turn, the proper method of their measurement. So, general strategy of quantitative description of fractal objects, in general case, should contain the following direct or reverse procedure: the choice of measure - the set of appropriate probabilities - the measuring method of these probabilities.

We choose the reverse procedure. So long as in the present work the averaged masses of elemental components of atmospheric PM-multifractal are measured, the geometrical probabilities, which can be constructed by experimental data for some fixed $\varepsilon$, have the practically unambiguous form:

\begin{equation}
\label{eq25}
p_i \left( \varepsilon = const \right) = \frac{m_i / \rho_i}{\sum\limits_i m_i/ \rho_i} =  \frac{C_i / \rho_i}{\sum\limits_i C_i/ \rho_i},
\end{equation}

\noindent where $\rho_i$ is the specific gravity of the $i$-th component of secondary aerosol.

Since the random nature of atmospheric PM formation is a priori determined by the random process of multicomponent diffusion-limited aggregation (DLA), we used the so-called harmonic measure \cite{ref-18} to describe quantitatively a stochastic surface inhomogeneity or, more precisely, to study an evolution of possible growth of cluster PM diameter. 

In practice, a harmonic measure is estimated in the following way. Because the perimeter of clusters, which form due to DLA, is proportional to their mass, the number of knots $N_p$ on the perimeter, i.e., the number of possible growing-points, is proportional to the number of cells $N$ in a cluster. Both these magnitudes, $N_p$ and $N$, change according to the power law (\ref{eq5}) depending on the cluster diameter $L$. From here it follows that all the knots $N_p$, which belong to the perimeter of such clusters, have a nonzero probability what a randomly wandering particle will turn out in them, i.e., they are the carriers of harmonic measure $M_d(q, \varepsilon_L)$:

\begin{equation}
M_d \left( q, \varepsilon_L \right) = \sum\limits_{i=1}^{N_p} p_i^q \cdot \left( \frac{\varepsilon}{L} \right)^d = Z \left( q, \varepsilon_L \right) \cdot \varepsilon_L^d  \xrightarrow[L\rightarrow \infty]{} \begin{cases} 0, & d > \tau(q) \\ \infty, & d < \tau(q) \end{cases},
\end{equation}

\noindent where $Z(q,L_\varepsilon)$ is the generalized statistical sum in the interval $ -\infty < q < \infty$, $\tau(q)$ is the index of mass, at which a measure does not become zero or infinity at $L \rightarrow \infty (\varepsilon_L \rightarrow 0)$.

It is obvious, that in such a form the harmonic measure is described by the full index sequence $\tau(q)$, which determines according to what power law the probabilities $p_i$ change depending on $L$. At the same time, the spectrum of fractal dimensions for the harmonic measure is calculated in the usual way, but using the “Brownian particles-probes” of fixed diameter $\varepsilon$ for study of possible growth of the cluster diameter $L$. From (26) it follows that in this case the generalized statistical sum $Z(q, \varepsilon_L)$ can be represented in the form

\begin{equation}
\label{eq27}
Z \left( q, \varepsilon_L \right) = \sum_{i=1}^{N_p} p_i^q \sim \varepsilon_L^{- \tau(q)}.
\end{equation}

As is known from numerical simulation of a harmonic measure, when the DLA cluster surface is probed by the large number of randomly wandering particles, the peaks of “high” asperities in such a fractal aggregate have greater possibilities than the peaks of “low” asperities. So, if possible growing-points on the perimeter of our aerosol multifractal to renumber by the index $i=1,\ldots, N_p,$ the set of probabilities

\begin{equation}
\label{eq28}
\Re = \left\lbrace p_i \right\rbrace _{i=1}^{N_p},
\end{equation}

\noindent composed of the probabilities of Eq. (\ref{eq25}) type will emulate the possible set of interaction cross-sections between Brownian particle and atmospheric PM-multifractal surface, which consists of the $N_p$ groups of identical atoms distributed on the surface. Each of these groups characterizes the $i$-th elemental component in the one atmospheric PM.

A situation is intensified by the fact that by virtue of (\ref{eq17}) each of the independent components obeys the Gauss distribution which, as is known \cite{ref-22}, belong to the class of infinitely divisible distributions or, more specifically, to the class of so-called $\alpha$-stable distributions. This means that although the Gauss distribution has different parameters (the average $\mu_i = \left\langle \ln{p_i} \right\rangle$ and variance $\sigma_i^2= \mbox{var}(\ln{p_i})$ for each of components, the final distribution is the Gauss distribution too, but with the parameters $ \mu = \left\langle \ln{p} \right\rangle = \sum \mu_i$  and $\sigma^2 = \sum \sigma_i^2$. From here it follows that the parameters of the two-dimensional normal distribution of all corresponding components in the plane $ \left\lbrace p_1, p_2 \right\rbrace$ are connected by the equations of direct linear regression:

\begin{equation}
\label{eq29}
\ln{p_{1}} - \left\langle \ln{p_{1}} \right\rangle = r \frac{\sigma_{1}}{\sigma_{2}} \left[ \ln{p_{2}} - \left\langle \ln{p_{2}} \right\rangle \right]
\end{equation}

\noindent and inverse linear regression:

\begin{equation}
\ln{p_{2}} - \left\langle \ln{p_{2}} \right\rangle = r \frac{\sigma_{2}}{\sigma_{1}} \left[ \ln{p_{1}} - \left\langle \ln{p_{1}} \right\rangle \right].
\end{equation}

So, validity of the assumption (\ref{eq28}) for the secondary aerosol or, that is the same, validity of the application of harmonic measure for the quantitative description of aerosol stochastic fractal surface, will be proven if simultaneous consideration of (\ref{eq28}) and equation of direct linear regression (\ref{eq29}) will result in an equation identical to the equation of direct linear regression (\ref{eq3}).

Therefore, taking into account (\ref{eq28}) we write down, for example, the equation of direct linear regression or, in other words, the condition of linear correlation between the samples of $i$-th component concentrations $ \left\lbrace \ln{ \left( C_{1i} / \rho_i \right)} \right\rbrace$ and $ \left\lbrace \ln{ \left( C_{2i} / \rho_i \right)} \right\rbrace$ in an atmospheric aerosol measured in different places (indexes 1 and 2):

\begin{equation}
\ln{ \left( C_{1i} / \rho_i \right)} = a_{12} + b_{12} \ln{ \left( C_{2i} / \rho_i \right)}, ~~~ a_{12} = \frac{1}{N_p} \ln{\frac{\prod\limits_{i=1}^{N_p} \left( C_{1i} / \rho_i \right)}{ \left( \prod\limits_{i=1}^{N_p} \left( C_{2i} / \rho_i \right) \right) ^{b_{12}}}}, ~~~ b_{12} = r \frac{\sigma_1}{\sigma_2}.
\end{equation}

It is obvious, that this equation completely coincides with the equation of linear regression (\ref{eq3}), but is theoretically obtained on basis of the Gauss distribution of the random magnitude $\ln{p_i}$ and not in an empirical way.

Physical interpretation of the intercept $a_{12}$ is evident from the expression (\ref{eq29}), whereas meaning of the regression coefficient $b_{12}$ becomes clear, according to (\ref{eq19}) and (\ref{eq29}), from the following expression: 

\begin{equation}
b_{12} = r \left( \frac{\mbox{var} \left( \ln{p_1} \right)}{\mbox{var} \left( \ln{p_2} \right)} \right)^{1/2} = r \left[ \frac{\ln{ \left( L_1 / \varepsilon_1 \right)}}{\ln{ \left( L_2 / \varepsilon_2 \right)}} \right]^{1/2},
\end{equation}

\noindent where $L_1$ and $L_2$ are the average sizes of separate atmospheric PM-multifractals typical for the atmosphere of investigated regions (indexes 1 and 2) of the Earth, $\varepsilon_1$ and $\varepsilon_2$ are the cell sizes into which the corresponding atmospheric PM-multifractals are divided.

Below we give a computational procedure algorithm for identification of the generalized fractal dimension $D_q$ spectra and function $ f(\alpha)$. It is obvious, that such a problem can be solved by the following redundant system of nonlinear equations of Eqs. (\ref{eq15}), (\ref{eq16}), (\ref{eq18}) and (\ref{eq19}) type: 

\begin{equation}
\label{eq33}
\begin{array}{c}
\ln{ \left\langle p \right\rangle} = \left( 2 \alpha_0 - 3 D_0 \right) \ln{ \left( L / \varepsilon \right)}, \\
\\
\mbox{var}(p) = \left( L / \varepsilon \right)^{2 \left( 2 \alpha_0 - 3 D_0 \right)} \left[ \left( L / \varepsilon \right)^{2 \left( 2 \alpha_0 - D_0 \right)} -1 \right].
\end{array}
\end{equation}

\begin{equation}
\label{eq34}
\begin{array}{c}
\left\langle \ln{ p } \right\rangle = \mu + \sigma^2 = \left( \alpha_0 - 2 D_0 \right) \ln{ \left( L / \varepsilon \right)}, \\
\\
\mbox{var} \left( \ln{p} \right) = \sigma^2 = 2 \left( \alpha_0 - D_0 \right) \ln{ \left( L / \varepsilon \right)},
\end{array}
\end{equation}

\noindent where $\varepsilon$ is the cubic cell fixed size, into which the bounded region £ of size $L$ in Euclidian space of dimension $d$ is divided.

To solve the system of equations (\ref{eq33})-(\ref{eq34}) with respect to the variables $\alpha_0$, $D_0$ and $\varepsilon$, $\varepsilon_L$ it is necessary and sufficient to measure experimentally the $i$-th components of the concentration sample $ \left\lbrace C_i \right\rbrace$ in the unit volume of atmospheric air (see section 3) and size distribution of atmospheric PM for determination of the average size $L$.

It will be recalled that from the physical standpoint so-called the box counting dimension $D_0$, the entropy dimension $D_1$ and the correlation dimension $D_2$ are the most interesting in the spectrum of the generalized fractal dimensions $D_q$ corresponding to different multifractal inhomogeneities. Within the framework of the notions and definitions of multifractal theory mentioned above we describe below the simple procedure for finding of spectrum of the generalized fractal dimensions $D_q$, taking into account the solution of the system of equations (\ref{eq33})-(\ref{eq34}).

From (\ref{eq27}) it follows that in our case a multifractal is characterized by the nonlinear function $\tau(q)$ of moments $q$

\begin{equation}
\label{eq35}
 \tau (q) = \lim_{\varepsilon_L \rightarrow 0} \frac{\ln{Z \left( q, \varepsilon_L \right)}}{ \ln{\varepsilon_L} }.
\end{equation}

As well as before we consider a fractal object, which occupies some bounded region £ of “running” size $L$ (so that $\varepsilon_L \rightarrow 0$) in Euclidian space of dimension $d$. Then spectrum of the generalized fractal dimensions $D_q$ characterizing the multifractal statistical inhomogeneity (the distribution of points in the region £) is determined by the relation

\begin{equation}
\label{eq36}
D_q = \frac{\tau (q)}{q - 1},
\end{equation}

\noindent where $\left( q - 1 \right)$ is the numerical factor, which normalizes the function $\tau(q)$ so that the equality $D_q = d$ is fulfilled for a set of constant density in the $d$-dimensional Euclidian space.

Further, we are interested in the known in theory of multifractal connection between the mass index $\tau(q)$ and the multifractal function $f(\alpha)$ by which the spectrum of generalized fractal dimensions $D_q$ is determined

\begin{equation}
\label{37}
D_q = \frac{\tau(q)}{q - 1} = \frac{1}{q - 1} \left[ q \cdot a(q) - f \left( a (q) \right) \right].
\end{equation}

It is obvious, that in our case, when the sample $ \left\lbrace p_i \right\rbrace $ is experimentally determined and the cell size $ \varepsilon << L$ is numerically evaluated (by the system of equations (\ref{eq33})-(\ref{eq34})), the spectrum of generalized fractal dimensions $D_q$ (\ref{eq36}) can be obtained by the expression for the mass index $\tau(q)$ (\ref{eq35}):

\begin{equation}
\label{eq38}
D_q = \frac{\tau(q)}{q - 1} \approx \frac{\ln{\sum\limits_{i=1}^{N_p} p_i^q }}{(q-1) \ln{\varepsilon_L}}.
\end{equation}

Finally, joint using of the Legendre transformation

\begin{equation}
\label{eq39}
\alpha = \frac{d \tau}{d q},
\end{equation}

\begin{equation}
f (\alpha) = q \frac{d \tau}{d q} - \tau,
\end{equation}

\noindent which sets direct algorithm for transition from the variables $\left\lbrace q,\tau(q) \right\rbrace$ to the variables $ \left\lbrace \alpha, f(\alpha) \right\rbrace$, and the approximate analytical expression (\ref{eq38}) for the function $D_q$ makes it possible to determine  an expression for the multifractal function $f(\alpha)$.  

Now we will consider the special case of search of the box counting dimension $D_0$ and the entropy dimension $D_1$. One of goals of this consideration is the validation of assumption of smallness of the magnitude $D''_{q=0}$ in the expression (\ref{eq10}), which was used for the derivation of log normal distribution of the random magnitude $p_i$ (\ref{eq13}).

It is easy to show that combined using of Eq. (\ref{eq9}) and the inverse Legendre transformation sets, which sets transition from the variables $ \left\lbrace \alpha , f(\alpha) \right\rbrace$ to the variables $ \left\lbrace q,\tau(q) \right\rbrace$, gives the following dependence of $\alpha(q)$ on the moments $q$: 

\begin{equation}
\label{eq41}
\alpha (q)  = 2 q \left( \alpha_0 - D_0 \right) + \alpha_0.
\end{equation}

Substituting (\ref{eq41}) and (\ref{eq9}) into (\ref{37}), we obtain the approximate expression for the spectrum of generalized fractal dimensions $D_q (q =0.1)$ depending on $\alpha_0$ and $D_0$:

\begin{equation}
D_q = \frac{1}{q - 1} \left[ q^2 \left(\alpha_0 - D_0 \right) + q \alpha_0 - D_0 \right].
\end{equation}

Thus, we can write down the expressions for the box counting dimension $D_0$ and the entropy dimension $D_1$ depending on $\alpha_0$ and $D_0$

\begin{equation}
D_0 = D_0,
\end{equation}

\begin{equation}
\label{eq44}
D_1 = \lim_{q \rightarrow 1} \frac{q \cdot a(q) - f \left( a(q) \right)}{q -1} = 2D_0 - \alpha_0.
\end{equation}

Here it is necessary to make a few remarks. It will be recalled that $ f(\alpha_{q=1}) = D_1$ is  the value of fractal dimension of that subset of the region £, which makes a most contribution to the statistical sum (\ref{eq36}) at $q=1$. However, by virtue of normalizing condition the statistical sum (\ref{eq36}) is equal to unity at $q=1$ and does not depend on the cell size $\varepsilon$, on which the region £ is divided. Thus, this most contribution also is of order unity. Therefore, in this case (and only in this case!) the probabilities of cell occupation $p_i \approx \varepsilon_L^\alpha$ (\ref{eq6}) are inversely proportional to the total number of cells $n(\varepsilon) \approx \varepsilon_L^{-f(\alpha)}$ (\ref{eq7}), i.e., the condition $f(\alpha) = \alpha$ is fulfilled.

So, the parameters of system of the equations (\ref{eq33})-(\ref{eq34}) obtained by the expression (\ref{eq9}) can not in essence contain information about the generalized fractal dimensions $D_q$ for absolute value of the moments $q$ greater than unity (i.e., $\vert q \vert > 1$).

\begin{figure}
\centering
\includegraphics[width=90mm]{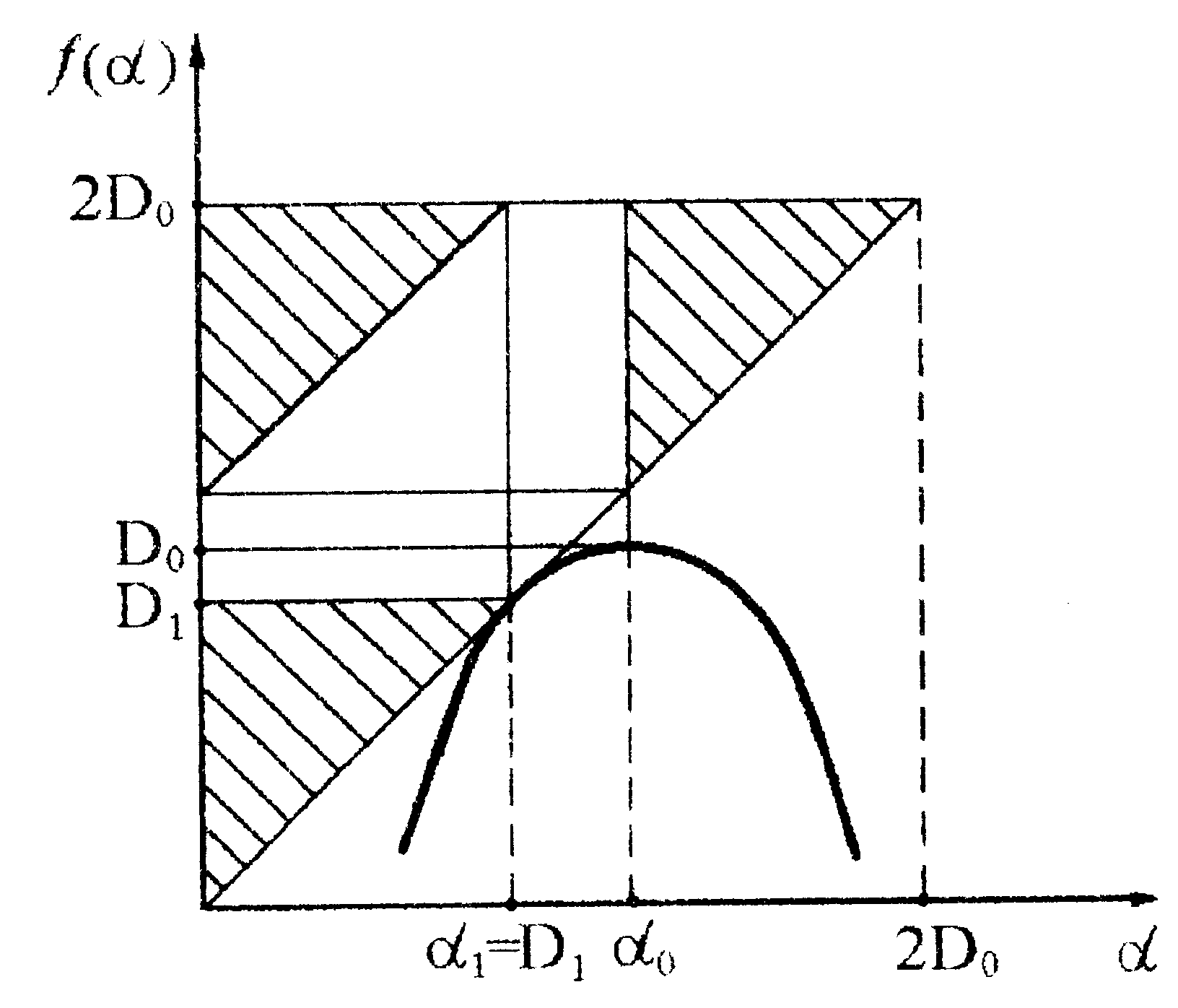} 
\caption{Geometrical method for determination of the entropy dimension $D_1$, which leads to the obvious equality $D_1=2D_0 - \alpha_0$.}
\label{fig8}
\end{figure}

Secondly, it is easy to show that the expression (\ref{eq39}) for the entropy dimension $D_1$ does not depend on concrete type of the function $f(\alpha)$, but is determined by its properties, for example, by symmetry $f(\alpha) = \alpha$, $f'(\alpha) = 1$ and convexity $f''(\alpha) > 0$. The geometrical method for determination of the entropy dimension $D_1$ shown in Figure \ref{fig8} is simultaneously the geometrical proof of assertion (\ref{eq44}).

Thirdly, the expressions for the entropy dimension $D_1$ obtained by parabolic approximation of the function $f(\alpha)$ and geometrical method (Figure \ref{fig8}) are equivalent. This means that the magnitude $D''_{q=0}$ in the expression (\ref{eq10}) is equal to zero. Thus, our assumption of smallness of the magnitude $D''_{q=0}$ in the expression (\ref{eq10}) is mathematically valid.

In the end, we note that the knowledge of generalized fractal dimensions $D_q$, the correlation dimension $D_2$ and especially $D_1$, which describes an information loss rate during multifractal dynamic evolution, plays the key role for an understanding of the mechanism of secondary aerosol formation, since makes it possible to simulate a scaling structure of an atmospheric PM with well-defined typical scales. Returning to the initial problem of distribution of points over the fractal set £, it is possible to say that the magnitude $D_1$ gives an information necessary for determination of point location in some cell, while the correlation dimension $D_2$ determines the probability what a distance between the two randomly chosen points is less than $\varepsilon_L$. In other words, when the relative cell size tends to zero ($\varepsilon_L \rightarrow 0$), these magnitudes are anticorrelated, i.e. the entropy $D_1$ decreases, while the multifractal correlation function $D_2$ increases.


\section{Conclusions}

Comparative analysis of different pairs of experimental normalized concentration values of  atmospheric PM elements measured in different regions of the Earth shows a stable linear (on a logarithmic scale) correlation ($r=1$) dependence on different time scales (from average daily to annual). That points to a power law increase of every atmospheric PM element mass (volume) and simultaneously to the cause of this increase -- the fractal nature of the genesis of atmospheric PM. 

Within the framework of multifractal geometry it is shown that the mass (volume) distribution of the atmospheric PM elemental components is a log normal distribution, which on the logarithmic scale with respect to the random variable (elemental component mass) is identical to the normal distribution. This means that the parameters of the two-dimensional normal distribution with respect to corresponding atmospheric PM-multifractal elemental components, which are measured in different regions, are a priory connected by equations of direct and inverse linear regression, and the experimental manifestation of this fact is the linear (on a logarithmic scale) correlation between the concentrations of the same elemental components in different sets of experimental atmospheric PM data.

We would like to note here that a degree of our understanding of the mechanism of atmospheric PM formation, which due to aggregation on inoculating centres (primary aerosols ($D_p < 1\mu m$)) show a scaling structure with well-defined typical scales, can be described by the known phrase: “ …we do not know till now why clusters become fractals, however we begin to understand how their fractal structure is realized, and how their fractal dimension is related to the physical process” \cite{ref-23}. This made it possible to show that the spectrum of fractal dimensions of multifractal, which is a multicomponent (by elements) aerosol, always predetermines the log normal type of statistics or, in other words, the log normal type of mass (volume) distribution of the i-th component of atmospheric PM.

It is theoretically shown, how solving the system of nonlinear equations composed of the first moments (the average and variance) of a log normal and normal distributions, it is possible to determine the multifractal function $f(\alpha)$ and spectrum of fractal dimensions $D_q$ for separate averaged atmospheric PM, which are the global characteristics of genesis of atmospheric PM and does not depend on the local place of registration (measurement).

We should note here that the results of this work allow an approach to formulation of the very important problem of aerosol dynamics and its implications for global aerosol climatology, which is connected with the global atmospheric circulation and the life cycle of troposphere aerosols \cite{ref-3, ref-21}. It is known that absorption by the Earth's solar short-wave radiation at the given point is not compensated by outgoing long-wave radiation, although the integral heat balance is constant. This constant is supported by transfer of excess tropical heat energy to high-latitude regions by the aid of natural oceanic and atmospheric transport, which provides the stable heat regime of the Earth. It is evident that using data about elemental and dispersed atmospheric PM composition in different regions of the Earth which are “broader-based” than today, one can create the map of latitude atmospheric PM mass and size distribution. This would allow an analysis of the interconnection between processes of ocean-atmosphere circulation and atmospheric PM genesis through the surprising ability of atmospheric PM for long range transfer, in spite of its short “lifetime” (about 10 days) in the troposphere. If also to take into consideration the evident possibility of determination of latitude inoculating centers (i.e., primary aerosol) distribution, this can lead to a deeper understanding of the details of aerosol formation and evolution, since the natural heat and dynamic oscillations of the global ocean and atmosphere are quite significant and should impact influence primary aerosol formation dynamics and fractal genesis of secondary atmospheric aerosol, respectively.

It is important to note also that continuous monitoring of the main characteristics of  South Pole aerosols as a standard of relatively pure air, and the aerosols of large cities, which are powerful sources of anthropogenic pollution, allows to determine the change of chemical and dispersed compositions of aerosol pollution. Such data are necessary for a scientifically-founded health evaluation of environmental quality, as well as for the planning and development of an air pollution decrease strategy in cities.


\bibliographystyle{mdpi}
\makeatletter
\renewcommand\@biblabel[1]{#1. }
\makeatother

\begin{thebibliography}{1}

\bibitem{ref-1}
Maenhaut, W.; Zoller, W.H. Determination of the chemical composition of the South Pole aerosol by instrumental neutron activation analysis. {\em J. Radioanal. Chem.} {\bf 1977}, {\em 37}, 637-650.

\bibitem{ref-2}
Pushkin, S.G.; Mihaylov, V.A. {\em Comparative Neutron Activation Analysis: Study of Atmospheric Aerosols}; Nauka, Siberian Department: Novosibirsk, 1989.

\bibitem{ref-3}
Raes, F.; van Dingenen, R.; Vignati, E.; Wilson, J.; Putaud, J.P.; Seinfeld J.H.; Adams, P.  Formation and cycling of aerosols in the global troposphere. {\em Atmos. Environ}. {\bf 2000}, {\em 34}, 4215-4240.

\bibitem{ref-4}
Rusov, V.D.; Glushkov, A.V.; Vaschenko, V.N. {\em Astrophysical Model of the Earth Global Climate}; Naukova Dumka: Kiev, 2003 (in Russian).

\bibitem{ref-5}
Figen, Var; Yasushi Narita; Shigeru Tanaka. The concentration, trend and seasonal variation of metals in the atmosphere in 16 Japanese cities shown by the results of National Air Surveillance Network (NASN) from 1974 to 1996. {\em Atmos. Environ}. {\bf 2000}, {\em 34}, 2755-2770.

\bibitem{ref-6}
Brownlee, K.A. {\em  Statistical Theory and Methodology in Science and Engineering}; Ed. John Wiley \& Sons: New York, 1965.

\bibitem{ref-7}
Bendat, J.S.; Piersol, A.G. {\em Random Data: Analysis and Measurement Procedures}; Ed. John Wiley \& Sons: New York, 1986.

\bibitem{ref-8}
Schroeder, M.  Fractals, {\em  Chaos, Power Laws: Minutes from Infinite Paradise}; Ed. W. Freeman and Company: New York, 2000.

\bibitem{ref-9}
Witten, T.A.; Sander, L.M. Diffusion-limited aggregation: kinetic critical phenomenon. {\em Phys. Rev. Lett}.{\bf 1981}, {\em 47}, 1400-1403.

\bibitem{ref-10}
Zubarev, A.Yu.; Ivanov, A.O. Fractal structure of colloid aggregate. {\em Reports of Russian Academy of Sci}. 2002, 383, 472-477.

\bibitem{ref-11}
Maenhaut, W.; Francos, F.; Cafmeyer, J. The “Gent” Stacked Filter Unit (SFU) Sampler for Collection of Atmospheric Aerosols in Two Size Fractions: Description and Instructions for Installation and Use. Report No.NAHRES-19, IAEA: Vienna, 1993, pp. 249-263.

\bibitem{ref-12}
Hopke, P.K.; Hie, Y.; Raunemaa, T.; Biegalski, S.; Landsberger, S.; Maenhaut, W.; Artaxo, P.; Cohen, D. Characterization of the Gent stacked filter unit $PM_{10}$ Sampler. {\em Aerosol Sci. Tech}. {\bf 1997}, {\em 27}, 726-735.

\bibitem{ref-13}
Jaćimović, R.; Lazaru, A.; Mihajlović, D;, Ilić, R;, Stafilov, T. Determination of major and trace elements in some minerals by $k_0$-instrumental neutron activation analysis. {\em J. Radioanal. Nucl. Chem}. {\bf 2002}, 253, 427-434.

\bibitem{ref-14}
{\em HYPERMET-PC V5.0, User’s Manual}; Institute of Isotopes: Budapest, Hungary, 1997.

\bibitem{ref-15}
Kayzero/Solcoi$^{®}$ ver. 5a. {\em User’s Manual for Reactor-neutron Activation Analysis (NAA) Using the $k_0$-Standardization Method}; DSM Research: Geleen, Netherlands, 2003. 

\bibitem{ref-16}
Cronover, R.M. {\em Introduction to Fractals and Chaos}; Jones and Bartlett Publishers, 1995. 

\bibitem{ref-17}
Mandelbrot, B.B. {\em The fractal geometry of nature. Updated and Augmented}; W.H. Freeman and Company: New York, 2002. 

\bibitem{ref-18}
Feder, J. {\em Fractals}; Plenum Press: New York, 1988.

\bibitem{ref-19}
Bozhokin, S.V.; Parshin, D.A. {\em Fractals and Multifractals}; Scientific Publishing Centre "Regular and Chaotic Dynamics": Moscow-Izhevsk, 2001 (in Russian). 

\bibitem{ref-20}
Lai, F.S.; Friedlander, S.K.; Pich, J.; Hidy, G.M. The self-preserving particle size distribution for Brownian coagulation in the free-molecular regime. {\em J. Colloid Interf. Sci}. {\bf 1972}, {\em 39}, 395-405.

\bibitem{ref-21}
Raes, F.; Wilson, J.; van Dingenen, R. Aerosol dynamics and its implication for the global aerosol climatology. In {\em Aerosol Forcing of Climate}; Charson, R.J., Heintzenberg, J., Eds.; John Wiley \& Sons: New York, 1995.

\bibitem{ref-22}
Feller, W. {\em An Introduction to Probability Theory and its Applications}; John Wiley \& Sons: New York, 1971.

\bibitem{ref-23}
Bote, R.; Julen, R.; Kolb, M. Aggregation of Clusters. In {\em Fractals in Physics}; Pietronero, L., Tosatti, E., Eds.; North-Holland :Amsterdam, 1986, pp. 353-359.

\end{thebibliography}

\end{document}